\documentclass{arxivmanuscript_v2}

\usepackage{amsmath}
\usepackage{braket}
\usepackage{tabularx}
\usepackage{booktabs}
\usepackage{multirow}
\usepackage{comment}
\usepackage[
backend=biber,
style=numeric-comp,
sorting=none,
giveninits=true,
doi=true,
url=false
]{biblatex}

\renewbibmacro{in:}{}
\usepackage{makecell}
\usepackage{graphicx}%
\usepackage{multirow}%
\usepackage{amsmath,amssymb,amsfonts}%
\usepackage{amsthm}%
\usepackage{mathrsfs}%
\usepackage[title]{appendix}%
\usepackage{xcolor}%
\usepackage{textcomp}%
\usepackage{manyfoot}%
\usepackage{booktabs}%
\usepackage{algorithm}%
\usepackage{algorithmicx}%
\usepackage{algpseudocode}%
\usepackage{listings}%
\usepackage{braket}
\usepackage{dsfont}
\usepackage{subcaption}
\usepackage[normalem]{ulem}

\newcommand{\Vsrg}{V_{\text{SRG}}}

\begin{document}

\articletype{Article type} 

\title{Estimation of deuteron binding energy with renormalization group-based effective interactions using the variational quantum eigensolver}

\author{Sreelekshmi Pillai$^{1,2,*}$\orcid{0009-0001-0912-7564}, S. Ramanan$^{1,2}$\orcid{0000-0001-9316-874X}, V. Balakrishnan$^{2}$ \orcid{0000-0002-6050-9859} and S. Lakshmibala$^{2}$\orcid{0000-0001-9705-0845}}

\affil{$^1$Department of Physics, Indian Institute of Technology Madras, Chennai, India \looseness=-1}
\affil{$^2$Center for Quantum Information, Communication and Computing (CQuICC), Indian Institute of Technology Madras, Chennai, India \looseness=-1}

\affil{$^*$Author to whom any correspondence should be addressed.}

\email{sreelekshmi@physics.iitm.ac.in}

\keywords{Quantum computing for nuclei, Renormalization group, Deuteron, Chiral N4LO,  AV$_{18}$}

\begin{abstract}
We have obtained the energy  of the deuteron on a quantum simulator using the variational quantum eigensolver. We have 
employed realistic two-body
interactions, namely, chiral N4LO and AV$_{18}$, thus incorporating the role of tensor forces. These interactions are subsequently evolved to low resolution scales using the similarity 
renormalization group approach with parameter $\lambda$. The deuteron ground state energy  has been calculated in the truncated harmonic oscillator
basis, using the Qiskit-Aer simulator in both noise-free and noisy cases. The noise models have been taken from the actual IBM quantum hardware, and the results obtained have 
been extrapolated to the zero noise limit. The number of  harmonic oscillator basis states (hence qubits) needed for computing the energy to within 1 percent of the experimental value in the quantum 
simulator, decreases with decreasing {$\lambda$}.
We have analysed the extent of entanglement between 
oscillator modes using concurrence as the entanglement quantifier. It is seen that the entanglement decreases as $\lambda$ is lowered from the bare value to 
$\sim 1.0 \, \text{fm}^{-1}$ independent of the form of the bare interaction and the number of harmonic oscillator basis states.
\end{abstract}

\section{Introduction}
An effective description of a system of many nuclei starting from microscopic few body interactions poses challenges both conceptually as well as computationally. This is due to complexities at the level of the interactions, and the fact that these are strongly correlated systems. Hence, substantial research is still focused on  obtaining a comprehensive understanding of nuclear properties across the nuclear chart.
  
Progress in nuclear structure calculations can be often traced back to the availability of better computational resources or advances in the conceptual understanding of the few-body sector, or both. 
Further opportunities to handle numerically intensive problems more efficiently are being provided in recent years, owing to the 
tremendous effort invested in creating quantum computers
that promise to outperform classical computers. Therefore, for inherently numerically expensive problems such as those that arise in nuclear 
structure calculations, it is worthwhile to investigate how the quantum infrastructure can be utilized. Shell model calculations have been performed on quantum platforms for light and medium mass nuclei using phenomenological 
models and the ADAPT-VQE~\cite{Perez-Obiol2023,Bhoy2024}, which is a modified version of the variational quantum eigensolver (VQE) algorithm. Current research on nuclei using quantum computing platforms extends through a wide spectrum of topics such as computations of binding energy, investigations on resonances, neutron-proton pairing problems etc. (see for instance ~\cite{carrasco2026comparison,bacskan2026calculating,singh2025quantum,singh2025advancing,sarma2026low,zhang2025neutron,Siwach2021,Siwach2022,Perez-Obiol2023,Bhoy2024,Dumitrescu2018,Shehab2019,Gu2025,Sharma2024}).

 Quantification of nuclear correlations, primarily entanglement within modes and between spins of nuclear systems is currently being investigated~\cite{arumugam2025,bai2023spin,gu2023entanglement,johnson2023proton,karlsson2020quantum,kirchner2024entanglement,Kruppa2021,miller2023entanglement,perez2023quantum, Robin2021,brokemeier2025quantum}.
In particular, the 
deuteron ground state has provided an ideal framework for examining nuclear correlations \cite{arumugam2025, shilpashree2012spin}. The deuteron system has been studied on a quantum computer using realistic soft-core phenomenological Reid interaction~\cite{Siwach2021,Siwach2022}. Observables such as the binding energy and the quadrupole moment have been estimated (see, e.g.,~\cite{Siwach2022}). However, the results do not converge to realistically acceptable values for the qubit spaces considered. In particular, it has been emphasized in~\cite{Siwach2022} that large qubit spaces are required for convergence
to experimentally observed values for the deuteron observables (See Table~\ref{tab:litcompare} in Sec.~\ref{sect:results} for ready comparison with our results.).

Effective field theories (EFTs) and the renormalization group approach have helped in formulating effective interactions 
that are better starting points for perturbative calculations, compared to phenomenological interactions that conventionally have 
strong short-range repulsion. Therefore, for few-body systems, quantum computing platforms have been explored primarily 
within the framework of pionless EFT. For instance, the deuteron binding energy has been 
computed~\cite{Dumitrescu2018,Shehab2019,Gu2025},  
with merely two and three qubits, and by extrapolation to infinite dimensions. The results for the binding energy thus obtained and reported in the literature, are in agreement 
with the experiments to within 1$\%$. An extension to the $^3$He nucleus that includes three-body forces has been carried out 
in~\cite{Gu2025}, where the authors use a 3D spatial lattice and construct scalable nuclear Hamiltonians and assess the resource requirement on a quantum 
computing platform. More recently, the two-body scattering states have been computed using the VQE
algorithm in a harmonic oscillator (HO) basis and phase shifts have been extracted~\cite{Sharma2024}. In addition, LENPIC SCS-regulated N2LO interaction~\cite{epelbaum2019few} has been used
for the study of Gamow~-Teller nuclear beta decay transition amplitude, on a quantum simulator, using Okubo-Lee-Suzuki transformations with the nucleons confined in a harmonic trap~\cite{sarker2024calculation}.

In this paper we address the following two aspects. (1) We incorporate realistic 
interactions (in contrast to pionless EFT). Here the low and high-momentum modes are decoupled
using similarity renormalization group (SRG) transformation. This program is carried
out for the purpose of quantum computing, in particular, estimation of qubit requirement and
entanglement computation. (2) Examine the role
played by hardware noise and its subsequent mitigation, and obtain realistic values of observables in small 
qubit spaces. We note that a quantum simulator
with noise models from the actual hardware suffices for our study.

We use the chiral N4LO~\cite{Entem2017} and the Argonne V$_{18}$(AV$_{18}$)~\cite{av18} interactions 
and their SRG-evolved counterparts, 
to explore the connection between SRG-based interactions and the qubit requirement. We have computed the ground state energy of the deuteron (referred to as the deuteron energy, in the rest of the paper) using VQE. In 
particular, we have investigated how the numerical advantages available on a classical machine translate to the quantum context. 
Investigations on entanglement between oscillator modes can be crucial in quantum computing because they could reveal interesting links 
between single-particle states, efficient qubit mappings, ansatz design, and resource optimization for many-body nuclear simulations. 
Mode entanglement and correlations in two-nucleon systems have been analyzed earlier, for instance, in~\cite{Kruppa2021}. There have 
also been studies on the effect of the choice of basis on the entanglement between specific modes~\cite{Robin2021}. In this work 
we have explored mode entanglement and its dependence on the extent of renormalization as indicated by the SRG parameter $\lambda$. 

The organization of the paper is as follows. In Sec.~\ref{sect:formalism}, we briefly review the HO basis and its parameters, the VQE algorithm, the mapping of the deuteron Hamiltonian to the qubit basis, as well
as the interaction used for this calculation. The results and their dependence on the SRG evolution are presented in Sec.~\ref{sect:results}. Section~\ref{sect:concl} contains a summary of our investigations, and suggestions for possible extensions of this work. Details augmenting the main text are given
in the Appendices.
\section{Formalism}
\label{sect:formalism}

\subsection{Deuteron in the HO basis}
The $3$D isotropic HO basis $\{\ket{n, (l, S), J, M_J, T, M_T}\}$ is 
convenient for calculating the deuteron energy, treating it as an equivalent one-body problem on quantum computing platforms. 
Here $n$ is the principal quantum number, $J=1$ is the 
total angular momentum,  $S=1$ is the total spin, $T=0$ is the total isospin, $M_T=0$ is the third component 
of $T$ and $l$ is the relative orbital 
angular momentum which takes values $0$ and $2$ when the tensor force is included. The energy of an HO basis state for a given $n$, $l$ and oscillator frequency $\omega$ is $E_{n,l}=\hbar \omega \left( 2n+l+ \frac{3}{2}\right)$. The maximum value of $n$ in our computations is denoted by $n_{\text{max}}$. We define $N=n_{\text{max}}+1$ and therefore $n$ takes values $0,1,\dots N-1$. Thus for instance, 
for $N = 1$, $(n, l) = \{(0,0), (0,2) \}$. Since $l$ takes two values for each $n$, this results in $2N$ basis states, and therefore a $2N \times 2N$  dimensional Hamiltonian, with matrix elements
\begin{align}
&\braket{n, l|H|n', l'} 
= \nonumber \\ 
&\left(\frac{2}{\pi}\right)^2 \int_{0}^{\infty}  dk\, k^2\, \int_{0}^{\infty} dk'\, k'^2\, R_{nl}(k)\, R_{n'l'}(k')\braket{k, l|H|k', l'}.
\label{eq:ham_osc}
\end{align}
Here $R_{nl}(k)$ is the HO radial wavefunction in momentum space and $\braket{k, l|H|k', l'}$ are the Hamiltonian matrix 
elements in the momentum space partial wave basis. 
In Eq.~(\ref{eq:ham_osc}), the quantum numbers $S, J, M_J, T, M_T$ have been suppressed for notational
convenience. 

As a consequence of working with $2N$ oscillator basis states, ultraviolet (UV) and infrared (IR) cutoffs are naturally introduced, and the following has been established~\cite{Furnstahl2012}. In a truncated HO 
basis, it follows that the maximum energy of the oscillator $E_{\text{max}}=\hbar \omega \left( 2n_{\text{max}}+l_{\text{max}}+ \frac{3}{2}\right)$. Since $n_{\text{max}}=N-1$  and $l_{\text{max}}=2$, $E_\text{max}$ can be rewritten as $\hbar \omega \left(2 N + \frac{3}{2}\right)$. The necessary conditions for numerical convergence of the deuteron energy are: (1) $\lambda$ corresponding to the specific interaction considered should be smaller than the UV cutoff
\begin{align}
\lambda_{UV} = \sqrt{2 \left(2N + \frac{3}{2}\right)} \frac{1}{b},
\label{eq:conv_UV}
\end{align}
and (2) the nuclear radius should be smaller than
\begin{align}
L_0 = \sqrt{2 \left(2N + \frac{3}{2}\right)} b
\label{eq:conv_IR}
\end{align}
where $b = \sqrt{\tfrac{\hbar}{m\omega}}$ and $m$ is the reduced mass of the deuteron.
However, even if these two conditions are satisfied, the convergence of numerical results in the oscillator basis is not guaranteed because $\lambda$ is generally not a sharp cutoff, and the nuclear wavefunction could extend beyond the nuclear radius.

In the following, we have computed the deuteron energy using VQE. The salient features of this algorithm are described in the next section. 
\subsection{Variational Quantum Eigensolver} 
VQE is a hybrid quantum-classical algorithm used to compute the energy eigenvalues of quantum systems~\cite{Ayral2023,Cerezo2021}. 
In the second quantized form, the Hamiltonian in Eq.~(\ref{eq:ham_osc}) is given by
\begin{equation}
    {H}=\sum_{i}\sum_{j}\braket{i|T+V|j}a_j^{\dag} a_i.
    \label{hamiltonian}
\end{equation}
Here $a_j^{\dag}$, $a_j$ are the creation and annihilation operators of a deuteron in the HO basis $\{\ket{j}\}$. Each of the labels $i$ and $j$ denotes specific values of both $n$ 
and $l$, i.e., a mode. Using the Jordan-Wigner (JW) transformation~\cite{jordan1928paulische}, each mode $j$ can be mapped onto a qubit with occupation number $q_j$, as follows. Since Eq.~(\ref{hamiltonian}) is a one-body Hamiltonian, the JW mapping simplifies to $ a_j^{\dag}=\tfrac{1}{2}(X_j-iY_j)$, and  
$a_j=\tfrac{1}{2}(X_j+iY_j)$, with $X_j$ and $Y_j$ being the Pauli operators that act on the $j^{\rm th}$ qubit's space. The truncated HO basis is mapped onto the multiqubit basis denoted by $\{\ket{q_0, q_1, q_2, \ldots q_{2N-1}}\}$, with $N = n_{\text{max}} + 1$ and the occupation number $q_j =$ 1 or 0. For instance, in the four qubit case ($N = 2$), if the 0s state is occupied, this is mapped to the multiqubit state $\ket{1000}$. Similarly, if the 1s state is occupied, this is a multiqubit $\ket{0100}$. If the 0d or the 1d state is occupied, the multiqubit state is $\ket{0010}$ or $\ket{0001}$ respectively. It has been established in the literature that the JW mapping does not efficiently utilize the entire qubit Hilbert space, and that other mappings such as the Bravyi-Kitaev or the gray code encoding, do so more effectively~\cite{Seeley2012, Matteo2021, Siwach2021}. For our present purposes however, we have verified that the JW mapping suffices.

The parametrized state $\ket{\Psi(\boldsymbol{\theta})}$ required for VQE can be constructed using the unitary coupled-cluster (UCC) ansatz. It is given by
\begin{equation}\label{eq:ucc_ansatz}
    \ket{\Psi(\boldsymbol{\theta})} = U(\boldsymbol{\theta}) \ket{\phi_{\mathrm{HF}}},
\end{equation}
where $\ket{\phi_{\mathrm{HF}}}$ is the Hartree-Fock (HF) reference state, which in our case is $\ket{10000 \ldots}$, indicating that the lowest level ($0s$) is occupied.
The unitary operator $U(\boldsymbol{\theta})$, generated from the single excitation operators and the parameter vector
$\boldsymbol{\theta}$, is given by
\begin{equation}
    U(\boldsymbol{\theta}) = \exp\left[ \sum_{\alpha} {\theta}_{\alpha 0} \left( a_\alpha^\dagger a_0 - a_0^\dagger a_\alpha \right) \right],
    \label{eq:unitary_operator}
\end{equation}
where the index $\alpha$ denotes the unoccupied states, i.e, $\{\ket{j}\}$ where $j \neq 0$. 

The ansatz in Eq.~(\ref{eq:unitary_operator}) of Appendix~\ref{app/two qubit}, using single excitations, has to be represented as a quantum circuit for execution on a quantum computer. A two-qubit implementation of the UCC ansatz is given in Appendix~\ref{app/two qubit}.

The circuits obtained using the rules given in Appendix~\ref{app:cnot} can be further simplified by reducing the circuit depth, while still preserving their action on the reference state~\cite{Dumitrescu2018, Siwach2021}. 
Although reduction of circuit depth is important, we will demonstrate through our calculations that, for our current purposes, the CNOT staircase method suffices. 

The expectation value of the Hamiltonian is computed on a quantum simulator using the variational state. A classical optimizer iteratively updates the parameters until this value is minimized.

\subsection{Interaction}
\label{sect:interaction}

As mentioned earlier, we have carried out investigations with two different realistic forms of the two-body interaction, namely, the chiral N4LO~\cite{Entem2017} and the AV$_{18}$~\cite{av18} 
interactions evolved to low momenta using the SRG~\cite{Bogner2006} approach, to 
yield effective interactions referred to as $\Vsrg$, that depend on the renormalization scale $\lambda$. These have been shown to yield better numerical convergence compared to the 
phenomenological models, when used as inputs in many-body calculations~\cite{vlowk_review, sramanan_convergence}, due to the decoupling between low and high momentum states. 
The defining equation for the SRG evolution is
\begin{equation}
    \frac{dH(s)}{ds} = [\eta, H(s)]
\end{equation}
where $s^{-1/4} =  \lambda$ the  SRG resolution scale, $\eta$ is the generator and $H(s)$ is the evolved Hamiltonian. It is well known that working with Wilson's generator~\cite{Bogner2006} ($\eta = [T_\text{rel}, H(s)]$, and $T_\text{rel}$ is the kinetic energy
in the center of mass frame) leads to an approximate band diagonal form for the Hamiltonian, with $\lambda$ being the extent of the band.  This is shown in Fig.~\ref{fig:vsrg_n4lo_3s1} (Appendix~\ref{app:chiraln4lo}) for the chiral N4LO interaction. 
As the transformation is unitary, the two-body observables such as the deuteron energy and scattering phase shifts are independent of $\lambda$~\cite{Bogner2006,vlowk_review}.  
\begin{figure*}
    \includegraphics[width=\linewidth]{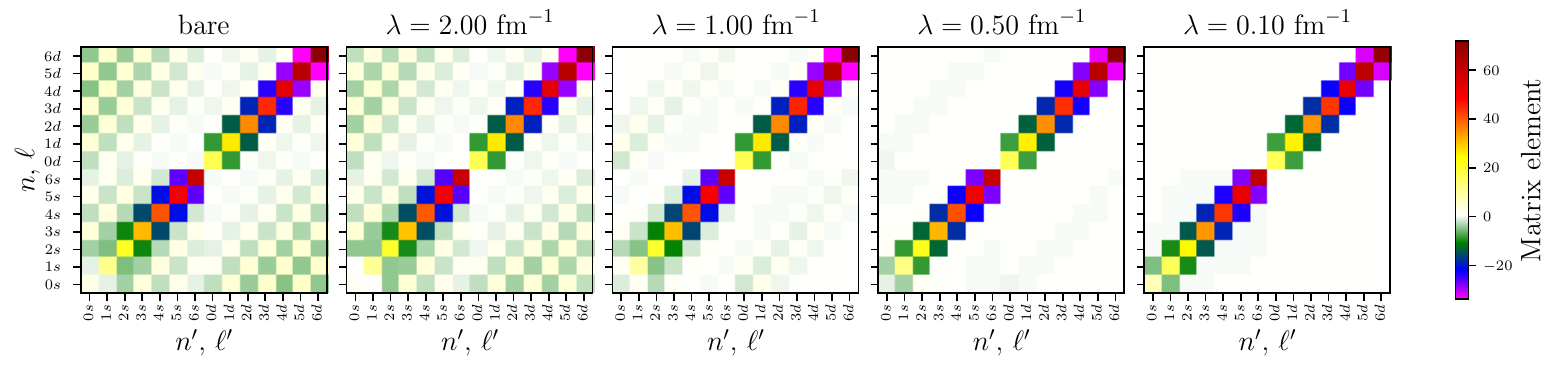}
    \caption{SRG evolution of chiral N4LO~\cite{Entem2017} in the $^3S_1-^3D_1$ channel in the HO basis ($N=7$, $\hbar\omega=9$ MeV). 
    }
    \label{fig:vsrg_n4lo_3s1_ho}
\end{figure*}
Using the momentum space partial wave matrix elements, the corresponding matrix elements in the HO basis have been obtained.
These are shown in Fig.~\ref{fig:vsrg_n4lo_3s1_ho} for different values of $\lambda$ for the chiral N4LO case. We see that for the bare interaction with $N = 7$,
all the oscillator states within our model space are coupled. However, as $\lambda$ decreases the coupling 
of the lower oscillator modes to the higher modes decreases. For a given $n$,
the interaction matrix elements between $l = 0$ and $2$ are strongly suppressed at lower values of $\lambda$, similar to the 
behavior of the momentum space matrix elements. This is due to the weakening of the repulsive tensor 
force~\cite{sramanan_convergence}. The matrix elements for AV$_{18}$ are similar to those corresponding to chiral N4LO for small values of $\lambda$ and are therefore not presented here. However, it is well known that the bare matrix elements of AV$_{18}$ have strong short-range repulsion. As we will demonstrate in the next section, this has implications for the convergence of the deuteron energy in small oscillator spaces.

In the next section, we present our results in the oscillator basis with $\Vsrg$ interactions derived 
from both chiral N4LO and AV$_{18}$ as inputs, using both exact diagonalization and the Qiskit-Aer simulator. 

\section{Results}
\label{sect:results} 
Using the effective interactions discussed in Sec.~\ref{sect:interaction}, we have computed the deuteron energy in the oscillator basis for the parameter ranges $0 < \hbar \omega \leqslant 40 \, \text{MeV}$, and $1 \leqslant N \leqslant 7$. We restrict ourselves to small values of $N$ as larger model spaces would imply higher qubit
demands. The SRG evolution being isospectral (energy eigenvalues are preserved) allows for very low values of $\lambda$. Therefore, we consider $0.1 \, \text{fm}^{-1} \leqslant \lambda \leqslant \lambda_\text{bare}$ ($\lambda_\text{bare} \sim 3.0 \, \text{fm}^{-1}$ for chiral N4LO and $\sim 30 \, \text{fm}^{-1}$ for AV$_{18}$). This computation was first carried out by exact diagonalization of the deuteron Hamiltonian in the HO basis, and subsequently using the VQE algorithm. The results of the former procedure serve as a check for the classical optimization which is an integral part of VQE.

\subsection{Exact diagonalization}
\label{subsect:conv}
\begin{figure*}
    \includegraphics[width=\linewidth]{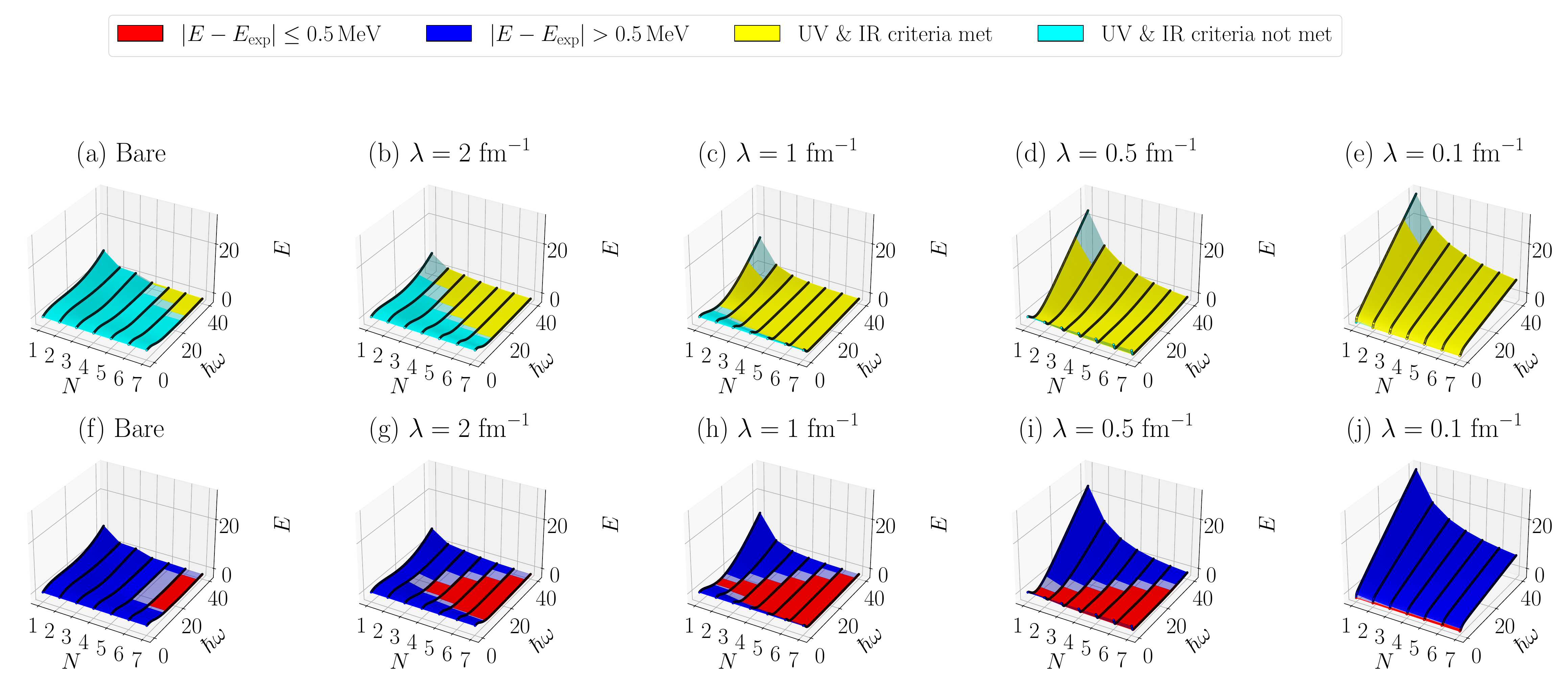}
    \caption{Deuteron energy as a function of $N$ and $\hbar \omega$ (in MeV) for the chiral N4LO interaction. Top panel (a) - (e): yellow region - IR and UV convergence criteria are met; cyan region - convergence criteria not met. Bottom panel (f) - (j): red region - deuteron energy converges ($|E - E_{\text{exp}}| \leqslant 0.5 \, \text{MeV})$; dark blue region - $|E - E_{\text{exp}}| > 0.5 \, \text{MeV}$.}  
    \label{fig:n4lo_conv}
\end{figure*}

{\color{blue}
\begin{figure*}
    \includegraphics[width=\linewidth]{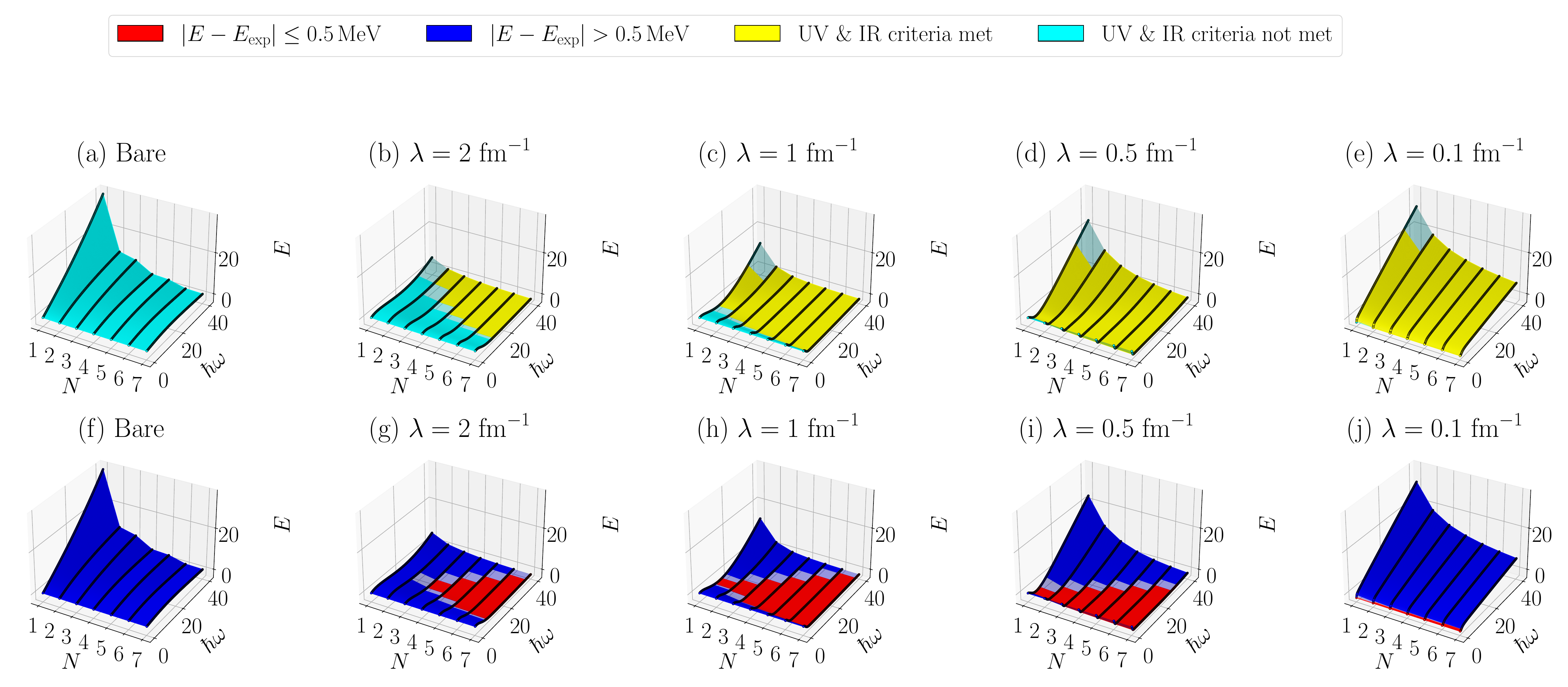}
     \caption{Deuteron energy as a function of $N$ and $\hbar \omega$ (in MeV) for the AV$_{18}$ interaction. As in Fig.~\ref{fig:n4lo_conv}, the oscillator parameters for which the IR and UV convergence criteria are satisfied (top panel) and the energy converges (bottom panel) are given by the shaded yellow region, and red region respectively. The regions
     where these convergence criteria are not satisfied are given by the cyan region (top panel) and the blue region (bottom panel). The behavior for small $\lambda$ is similar to that in Fig.~\ref{fig:n4lo_conv}. However, the bare AV$_{18}$ interaction does not converge in the entire parameter space that is explored.}  
    \label{fig:av18_conv}
\end{figure*}}

In Figs.~\ref{fig:n4lo_conv} and~\ref{fig:av18_conv}, the lowest energy eigenvalue obtained by exact diagonalization of the deuteron Hamiltonian in the HO basis is 
plotted as a function of $N$ and $\hbar \omega$ for different values of $\lambda$, for the chiral N4LO and AV$_{18}$ interactions respectively. In the top panel, in both Figs.~\ref{fig:n4lo_conv}  and~\ref{fig:av18_conv}
(a)-(e), the yellow (respectively cyan) region denotes the ranges of values of the oscillator parameters for which the convergence criteria (Eqs.~(\ref{eq:conv_UV}) and~(\ref{eq:conv_IR}) of 
Sec.~\ref{sect:formalism}) are satisfied (not satisfied). In the bottom panel (f)-(j), the red region 
marks the set of values of $N$ and $\hbar \omega$ for which the difference between the experimental and the computed values 
$|E - E_{\text{exp}}| \leqslant 0.5\, \text{MeV}$, while the difference exceeds 0.5 MeV in the dark blue region. As mentioned earlier, satisfying the convergence criteria alone does not guarantee 
that the deuteron energy will be close to the experimental value, as the position space wavefunction need not fall off sufficiently rapidly 
within the box of size $L_0$.

For the bare chiral N4LO interaction (Fig.~\ref{fig:n4lo_conv}), convergence in energy occurs for $N \sim 6$ and $15\, \text{MeV} < \hbar \omega < 40 \, \text{MeV}$. As $\lambda$ decreases, the set of oscillator parameters for which convergence is realized, increases until $\lambda \sim 0.5 \, \text{fm}^{-1}$ is reached. For $\lambda < 0.5 \, \text{fm}^{-1}$ however, convergence in energy is obtained even for $N = 1$, but for a narrower window of $\hbar \omega$ close to zero, implying that $L_0$ is relatively larger. Consequently the position wavefunction falls off to zero within the box. 

In contrast to chiral N4LO (Fig.~\ref{fig:n4lo_conv}), for AV$_{18}$ (Fig.~\ref{fig:av18_conv}) the bare interaction does not converge for the set of oscillator parameters considered. This 
is seen by the absence of the yellow shaded region in Fig.~\ref{fig:av18_conv}(a). As $\lambda$ decreases, converged results are obtained for a wide range of oscillator 
parameters, except at $\lambda = 0.1 \, \text{fm}^{-1}$, which supports only small values of $\hbar \omega$ but the entire range of values of $N$.

\subsection{Simulation with VQE}
\label{subsect:simulation}
We have used the Qiskit Software Development Kit (SDK) from IBM ~\cite{Javadi2024} for all our simulations. In particular, we have employed the Aer Estimator and Simulator. Since this is a
hybrid algorithm, the choice of classical optimizers (used for the ansatz parameter optimization) is crucial in determining the accuracy of the algorithm. We have used the COBYLA optimizer as 
it is gradient-free and works well in noisy environments. 

As a consequence of the JW mapping, the Hamiltonian is decomposed into a sum of Pauli strings. The expectation value of the Hamiltonian w.r.t. the variational state $\ket{\psi(\theta)}$, has been computed in two different ways: (1) using the statevector simulator and (2) with probabilistic 
simulations. The former models the exact circuit to yield results identical to those obtained from exact diagonalization, provided the 
classical optimization is reliable. In the probabilistic simulation, the quantum hardware performs measurements in the computational 
basis $\{|0\rangle,|1\rangle\}$, corresponding to a projective measurement of the 
Pauli-$Z$ observable on each qubit to produce a classical bit string. Each bit is mapped to a $Z$ eigenvalue ($0 \rightarrow +1$, 
$1 \rightarrow -1$), and the eigenvalue of a Pauli string is obtained as the product 
of the corresponding qubit eigenvalues. In the simulation, the entire circuit is 
evaluated $\sim 10^6$ times (`shots'). Averaging over the values obtained in all these shots 
the expectation value of the Hamiltonian is computed. This procedure gives rise to a sampling error even in the absence of hardware noise. Further, we have applied the Gaussian approximation to the output distribution. This replaces the discrete output distribution with a continuous Gaussian distribution with the same mean and variance. This procedure accelerates convergence while maintaining realistic gate noise 
characteristics. As a result of this choice, the readout errors have been suppressed, while the gate errors have been retained \cite{Javadi2024}.

We have performed 10 independent VQE runs for both the statevector and the probabilistic simulations. In the latter case, we have considered
both the presence and absence of hardware noise. For each VQE run, the initial set of parameters for the statevector simulation has been 
randomly chosen. For the probabilistic simulations, 
which needed 800,000 shots for convergence, we have used two types of initial parameters, namely, (1) the output 
parameters from the statevector simulator and (2) randomly generated parameters. The former set ensures that the given initial 
point is very close to the actual ground state wavefunction, thereby allowing for an investigation of the noise propagation 
on the quantum computing platform, while the latter set serves to test the convergence of the algorithm. The noise models 
from \texttt{ibm\_kyiv} and \texttt{ibm\_brisbane}\footnote{\texttt{ibm\_kyiv} and \texttt{ibm\_brisbane} have been retired. However, we have used the downloaded noise model while they were active.\label{myfootnote}}  
were used for this purpose. The circuit representing the ansatz may not match the hardware topology of the quantum computer (or its simulator). Hence, it is transpiled 
onto the chosen hardware. In order that the outliers do not contribute to our final result, we have computed the median and the Median Absolute Deviation (MAD) of the ground state 
energy~\cite{Siwach2021}.  

Since the two noise models considered give comparable results, we have only presented our results for \texttt{ibm\_brisbane} in 
Table~\ref{zne_table_srg_n4lo_brisbane} for the chiral N4LO and in Table~\ref{zne_table_srg_av18_split} for AV$_{18}$. In Table~\ref{zne_table_srg_n4lo_brisbane}, for specific values of $N$ and $\lambda$, the listed values of $\hbar \omega$  are those that yield the deuteron energy closest to the experimental value. 
We have verified that the results for the deuteron energy obtained from both the exact diagonalization and the statevector simulation (not shown in the table) are identical. Results from the simulations with shots excluding hardware noise are shown in column five. The 
deviation from the exact diagonalization result quantifies the sampling error as a function of the oscillator parameters and 
$\lambda$. In the sixth column, the results for the noisy simulation corresponding to input parameters obtained from 
the statevector simulator are shown, thereby providing an estimate of the gate errors. For $N = 2$, and $\lambda < 1.0 \, \text{fm}^{-1}$, the noisy outcomes are within 10$\%$ of the corresponding noise-free values. For $N = 3$ however, this error 
threshold is achieved only for the smallest value of $\lambda$ considered (i.e, $0.1 \, \text{fm}^{-1}$).
\begin{table*}[h]
\centering
\setlength{\tabcolsep}{5pt} 
\renewcommand{\arraystretch}{0.6}

\begin{tabular}{c c c c c c c c c}

\toprule
$N$ & $\lambda$ & $\hbar\omega$ & Exact diag. & Noiseless with shots & Noisy value & Poly-3 extrap. & random init. & \% error\\
\midrule
\multirow{5}{*}{1}
  & 2.00 & 18.6  & -0.1924 & -0.1880 & -0.1567 & -0.1811 & -0.1696 & $\approx6$  \\
  & 1.00 & 11.8  & -1.8497 & -1.8476 & -1.8336 & -1.8524 & -1.8500 & $\approx 0.1$ \\
  & 0.50 & 5.0   & -2.2112 & -2.2106 & -2.2004 & -2.2094 & -2.2082 & $\approx 0.05 $ \\
  & 0.10 & 0.1   & -2.1800 & -2.1800 & -2.1739 & -2.1764 & -2.1768 & $\approx 0.02$ \\
\midrule
\multirow{5}{*}{2}
  & 2.00 & 20.2  & -0.6440 & -0.6316 & -0.2205 & -0.6080 & -0.6945 & $\approx14$ \\
  & 1.00 & 11.7  & -1.8533 & -1.6121 & -1.8115 & -1.8043 & -1.8272 & $\approx1$ \\
  & 0.50 & 6.1   & -2.2137 & -2.2126 & -2.0754 & -2.1856 & -2.1999 & $\approx 0.6$ \\
  & 0.10 & 0.1   & -2.2250 & -2.2247 & -2.2012 & -2.2205 & -2.2220 & $\approx 0.06$ \\
\midrule
\multirow{5}{*}{3}
  & bare & 18.3  & -1.0508 & -1.0075 &  5.3427 & -0.8640 & -0.5826 & $\approx32$ \\
  & 2.00 & 15.8  & -1.8653 & -1.8392 &  3.6106 & -1.8079 & -1.6091 & $\approx11$ \\
  & 1.00 & 8.0   & -2.1964 & -2.1831 &  0.5241 & -2.1351 & -2.0600 & $\approx4$ \\
  & 0.50 & 3.0   & -2.2249 & -2.2221 & -1.1787 & -2.2278 & -2.1936 & $\approx2$ \\
  & 0.10 & 0.1   & -2.2251 & -2.2247 & -2.0567 & -2.2220 & -2.2159 & $\approx 0.2$ \\
\bottomrule

\end{tabular}
\caption{Zero-noise–extrapolated deuteron ground state energy for the SRG-evolved chiral N4LO interaction.
Noise model from \texttt{ibm\_brisbane}. For $N = 1$ and $2$, the bare interaction does not bind the deuteron for 
the range of $\hbar \omega$ considered and hence not listed.}
\label{zne_table_srg_n4lo_brisbane}
\end{table*}

\begin{table*}[h]
\centering
\setlength{\tabcolsep}{5pt}
\renewcommand{\arraystretch}{0.6}

\begin{tabular}{c c c c c c}

\toprule
$N$ & $\lambda$ & $\hbar\omega$ & Exact diag. & Statevector init. & random init. \\
\midrule
\multirow{3}{*}{1}
  & 1.0 & 11.8 & -1.8460 & -1.8464 & -1.8508 \\
  & 0.5 & 5.0  & -2.2142 & -2.2104 & -2.2119 \\
  & 0.1 & 0.1  & -2.1824 & -2.1794 & -2.1790 \\
\midrule
\multirow{3}{*}{2}
  & 1.0 & 11.8 & -1.8498 & -1.8322 & -1.8015 \\
  & 0.5 & 6.1  & -2.2167 & -2.1954 & -2.1966 \\
  & 0.1 & 0.1  & -2.2279 & -2.2233 & -2.2234 \\
\midrule
\multirow{4}{*}{3}
  & 2.0 & 17.3 & -1.8029 & -1.6723 & -1.6350 \\
  & 1.0 & 8.1  & -2.1983 & -2.1420 & -2.0240 \\
  & 0.5 & 6.1  & -2.2242 & -2.1953 &  1.7479 \\
  & 0.1 & 0.1  & -2.2280 & -2.2223 & -2.2196 \\
\bottomrule

\end{tabular}
\caption{Zero-noise–cubic-extrapolated deuteron ground state energy for the SRG-evolved AV$_{18}$ interaction.
Noise model from \texttt{ibm\_brisbane}. The bare interaction does not bind the deuteron for 
the range of $\hbar \omega$ and $N$ considered and hence not listed.}
\label{zne_table_srg_av18_split}
\end{table*}

\begin{figure}[h]
    \centering
    \includegraphics[width=\linewidth]{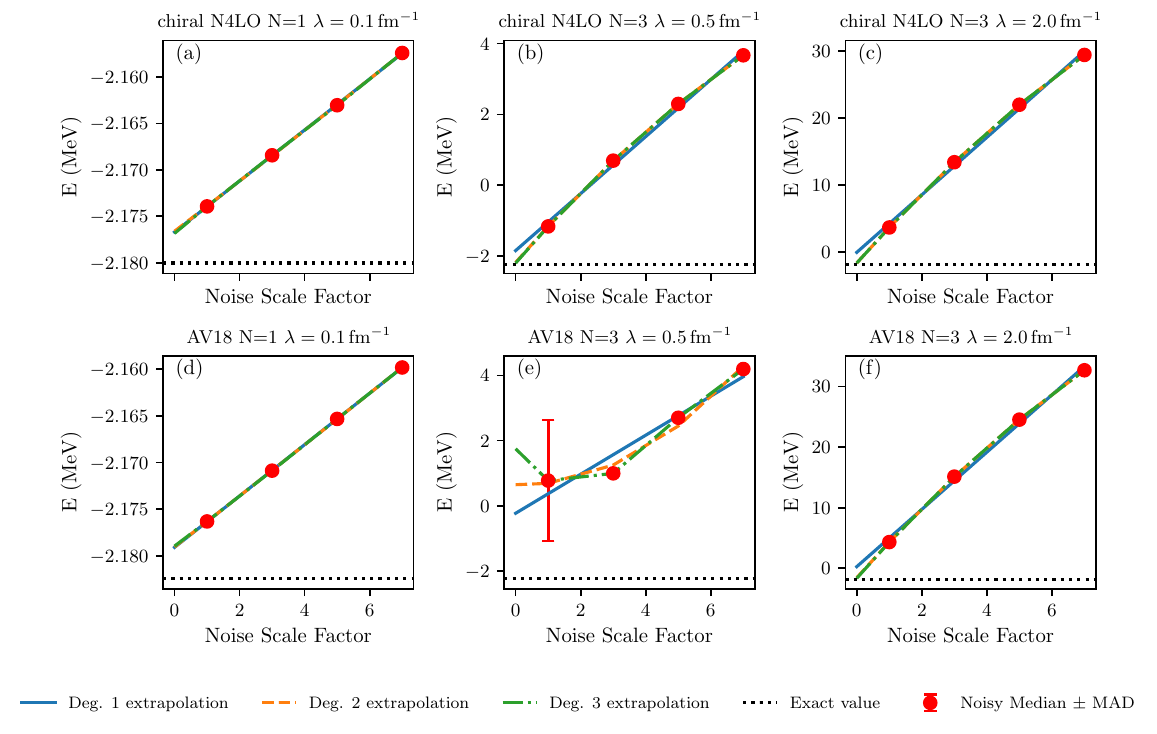}
    \caption{Zero-noise-extrapolated results obtained using \texttt{ibm\_{brisbane}} noise model and SRG-evolved interactions with random initial points. Chiral N4LO interaction with
    (a) $N=1,\ \lambda=1.0\,\mathrm{fm}^{-1}$ and $\hbar\omega=11.8$ MeV, 
    (b) $N=3,\ \lambda=0.5\,\mathrm{fm}^{-1}$ and $\hbar\omega=3.0$ MeV, 
    (c) $N=3,\ \lambda=2.0\,\mathrm{fm}^{-1}$ and $\hbar\omega=15.8$ MeV. AV$_{18}$ interaction with
    (d) $N=1,\ \lambda=1.0\,\mathrm{fm}^{-1}$ and $\hbar\omega=11.8$ MeV,
    (e) $N=3,\ \lambda=0.5\,\mathrm{fm}^{-1}$ and $\hbar\omega=3.3$ MeV, 
    (f) $N=3,\ \lambda=2.0\,\mathrm{fm}^{-1}$ and $\hbar\omega=17.3$ MeV.}
    
    \label{fig:znerep}
\end{figure}
 
We have attempted to mitigate the noise in our results using the zero-noise extrapolation (ZNE) with global circuit folding, which is a simple but effective method to reduce 
the effect of gate errors~\cite{circuitfolding}.
We have run the algorithm for different values of the noise scale factor and extrapolated it to zero noise using polynomial 
extrapolations of varying degree~\cite{Temme2017}. Figure~\ref{fig:znerep} illustrates examples of this implementation for both chiral N4LO and AV$_{18}$ with random initial points.  
We have observed that a cubic polynomial extrapolation is robust and
is generally sufficient across the parameter space considered. However, as
illustrated in Figs.~\ref{fig:znerep}(a) and (d), lower-degree polynomial extrapolations are often adequate for sufficiently small values of $N$, as the number of qubits and therefore the associated noise is less. Further, as seen from the figures, the dependence on the noise is often merely linear.  

We note that the initial points are randomly selected, and therefore statistical fluctuations can occasionally produce large variances in the fitted data, potentially leading 
to unreliable extrapolations (see for instance Fig.~\ref{fig:znerep}(e)). We have verified that in the majority of cases, a cubic polynomial extrapolation combined with 10 independent VQE runs yields stable and accurate zero-noise estimates. For the sake of methodological uniformity and consistency across all data sets, we therefore focus on the results obtained using cubic polynomial extrapolation throughout.

The results for ZNE with cubic polynomial extrapolation, for a range of oscillator parameters and $\lambda$ for the chiral N4LO interaction (Table~\ref{zne_table_srg_n4lo_brisbane} column seven) are presented for initial parameters obtained from the statevector simulator. The eighth column of Table~\ref{zne_table_srg_n4lo_brisbane} gives the final cubic polynomial extrapolated results when the initial parameters are 
chosen at random. The last column in Table~\ref{zne_table_srg_n4lo_brisbane} gives the relative error between the final results obtained for the two 
different initial inputs used for the noisy simulations, i.e., columns seven and eight of the table. We note that for any $N$,
decreasing $\lambda$ improves the convergence of the noisy simulation and for sufficiently small $N$ ($N = 1$ or 2), the errors
are about a percent or less for $\lambda \lesssim 1.0 \,\text{fm}^{-1}$.

Table~\ref{zne_table_srg_av18_split} shows the final cubic polynomial extrapolated results for the AV$_{18}$ interaction using input parameters obtained from both the statevector simulator as well as with random initial points. Once again, analogous to chiral N4LO, it is seen that for $\lambda \leq 1.0 \, \text{fm}^{-1}$, $N = 1$ (i.e., two qubits) gives reasonable results. In contrast, the bare AV$_{18}$ interaction does not give results that satisfy $|E-E_{\text{exp}}|\leq 0.5$ MeV, and hence are not shown in the table.

Our results demonstrate that the renormalized interactions require significantly fewer qubits for convergence to realistic values 
of the deuteron energy compared to the bare interactions, thereby providing enhanced mitigation of errors, as expected.

We now compare our results for the renormalized chiral N4LO and AV$_{18}$ interactions with existing literature (Table~\ref{tab:litcompare}). Our best results are for $\lambda < 1.0 \, \text{fm}^{-1}$. However, it may not always be feasible to lower $\lambda$ to such small values for systems with more than two nucleons, due to the growth of induced 
many-body forces. (We will elaborate further on this point in our concluding remarks.) 
We can therefore improve our results by correcting for the finite size of the HO basis, for this range of values of $\lambda$. 
One possibility is to use an oscillator variant of the L\"uscher's formula as in~\cite{Dumitrescu2018,Furnstahl2012}. 

It is evident from Table~\ref{tab:litcompare} that
REID68 estimates the deuteron energy to be 67.95 MeV and therefore fails to give any reasonable value. Further, the central interaction gives the ground state energy $\sim -1.0\, \text{MeV}$ even with four qubits. As is known from earlier results in the
literature, pionless EFT is a good candidate for producing realistic values of the deuteron energy with just two qubits. In our work we have shown that an alternative approach where also two qubits suffice, is 
provided by the SRG evolution of realistic interactions such as chiral N4LO and AV$_{18}$. 

\begin{table*}[h]
\centering
\begin{tabular}{|p{4.5cm} |p{2.2cm} |p{3cm} |p{2cm}|}

\hline
 & Interaction &  Deuteron energy (MeV)  &Qubits used\\
\hline

Dumitrescu et al. (2018)~\cite{Dumitrescu2018} &
Pionless EFT &
\makecell[l]{
$\approx -2.18^*$ \\
$\approx -2.28^*$
}  & \makecell[l]{
2\\
3
} \\
\hline

Shehab et al. (2019)~\cite{Shehab2019} &
Pionless EFT &
\makecell[l]{
$\approx -2.03$ \\
$\approx -2.22$ 
}  & 
\makecell[l]{
3\\
4
}  \\
\hline

Siwach et al. (2021)~\cite{Siwach2021} &
Pionless EFT &
\makecell[l]{
$\approx -2.18$ 
}  & \makecell[l]{
4
}\\
\hline

Siwach et al. (2021)~\cite{Siwach2021} &
Central &
\makecell[l]{
$\approx -1.028$
}  &  \makecell[l]{
4
} \\
\hline

Siwach et al. (2021)~\cite{Siwach2021, Siwach2022} &
REID68 &
\makecell[l]{
$\approx 67.95^{**}$\\
}  & \makecell[l]{
2\\
}\\
\hline

Present work&
chiral N4LO and AV$_{18}$&
\makecell[l]{
$\approx -2.2$ \\($\lambda=0.5\ \text{fm}^{-1}$)\\
$\approx -1.8$ to $-2.0$ \\($\lambda=1.0\ \text{fm}^{-1}$)\\
$\approx -1.6$ \\($\lambda=2.0\ \text{fm}^{-1}$)
}  & \makecell[l]{
2\\ \\
up to 6 \\ \\
6 
}\\\hline
\end{tabular}

\caption{Deuteron on a quantum computer: comparison between different approaches. $^*$ indicates L\"uscher's extrapolation~\cite{Dumitrescu2018,Furnstahl2012}. $^{**}$ indicates that the calculations were performed on QASM simulator without hardware noise.}
\label{tab:litcompare}
\end{table*}

\subsection{Mode entanglement}
\label{subsect:ent}
\begin{figure}
    \centering
    \includegraphics[width=\linewidth]{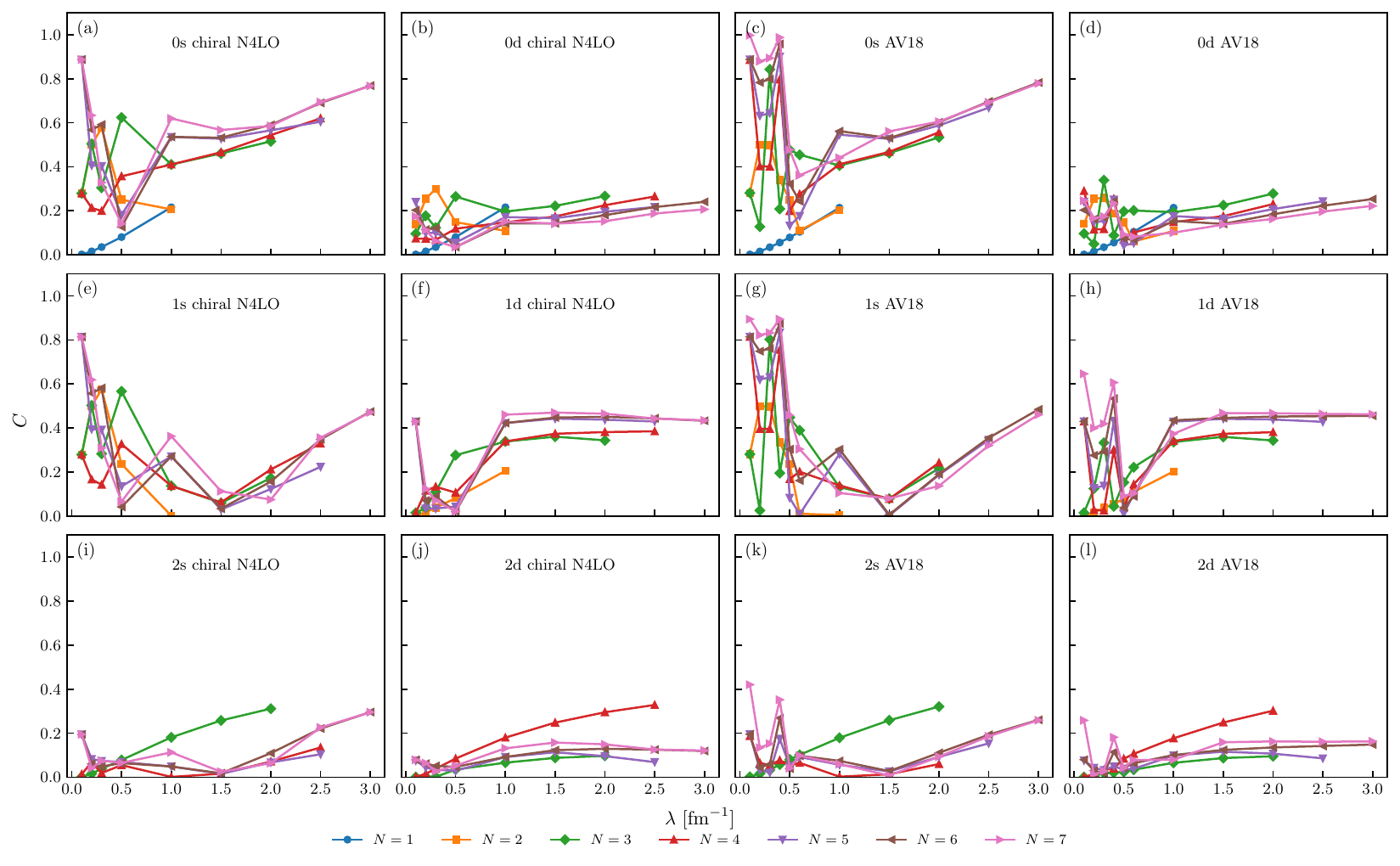}
   \caption{Single mode concurrence as a function of $\lambda$. 
(a) $0$s chiral N$4$LO, (b) $0$d chiral N$4$LO, (c) $0$s AV$_{18}$, (d) $0$d AV$_{18}$. 
(e) $1$s chiral N$4$LO, (f) $1$d chiral N$4$LO, (g) $1$s AV$_{18}$, (h) $1$d AV$_{18}$. 
(i) $2$s chiral N$4$LO, (j) $2$d chiral N$4$LO, (k) $2$s AV$_{18}$, (l) $2$d AV$_{18}$.  
For each $N$, $\hbar\omega$ is chosen to be the optimal one such that $|E - E_{\mathrm{exp}}| \le 0.5\,\mathrm{MeV}$. 
Modes absent when not allowed by the truncation: $1$s,$1$d for $N<2$; $2$s,$2$d for $N<3$.}
    \label{fig:conc_mode_wise}
\end{figure}

\begin{figure}[h!]
    \centering
    \includegraphics[width=\linewidth]{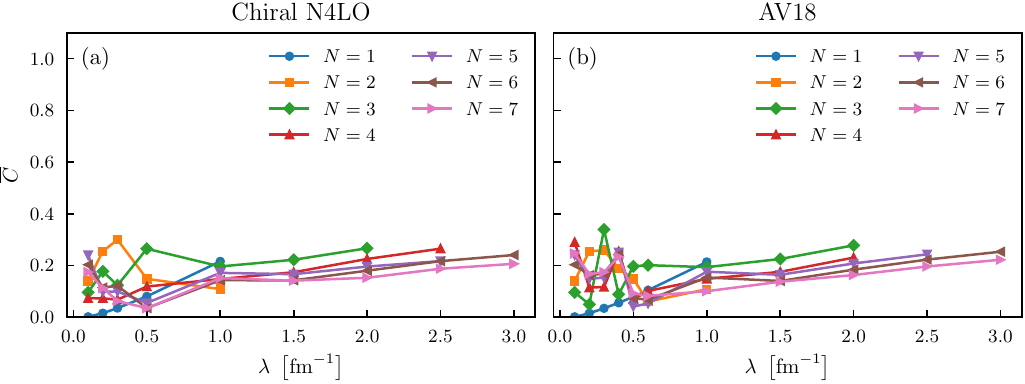}
    \caption{Average concurrence as a function of SRG parameter $\lambda$: 
    (a) chiral N4LO and
    (b)  AV$_{18}$. 
  The value chosen for $\hbar \omega$ is the optimal one 
such that  $|E - E_{\text{exp}}| \leqslant 0.5 \, \text{MeV}$}.
\label{fig:conc}

    \label{fig:conc_combined}
\end{figure}

In this section, we analyze the entanglement between the oscillator modes as a function of the oscillator parameters $N$ and 
$\hbar \omega$ for various values of $\lambda$.   
Bipartite mode entanglement can be quantified by distributing the total number of modes ($2N$) in different ways between two subsystems. We take a single mode to be the 
subsystem of interest, trace over the rest of the modes and use the concurrence
\begin{equation}
    C = \sqrt{2(1-(|\braket{\tilde{n}, \tilde{l}|\Psi}|^{4}+(1-|\braket{\tilde{n}, \tilde{l}|\Psi}|^{2})^{2}))}
    \label{eq:conc_sing_mode}
\end{equation}
to quantify bipartite entanglement. Here, $\tilde{n}$ and $\tilde{l}$ are the quantum numbers of the mode of interest and $\ket{\Psi}$ is the state of the deuteron (for details, see Appendix~\ref{app:mode_ent}). For $N \geq 2$, the single-mode concurrence is 
computed for all possible choices of the mode of interest and the results are presented in Fig.~\ref{fig:conc_mode_wise} for the first few modes for each $N$. The value
of $C$ as a function of $\lambda$ decreases as $\tilde{n}$ is increased from zero. For a given $\tilde{n}$, the $s$-mode ($\tilde{l} = 0$) is more entangled compared to the $d$-mode ($\tilde{l} = 2$). The concurrence for a given mode 
increases as $\lambda$ decreases till $\lambda \sim 0.5 \, \text{fm}^{-1}$. However, no reasonable systematics emerge for $\lambda < 0.5\, \text{fm}^{-1}$. In order to 
understand the general trends of the concurrence with $\lambda$ the results are averaged over this set to obtain the mean single-mode concurrence $\overline{C}$.
Figure~\ref{fig:conc_combined} shows $\overline{C}$ as a function of $\lambda$ for different values of $N$ for both chiral N4LO (Fig.~\ref{fig:conc}(a)) and AV$_{18}$ (Fig.~\ref{fig:conc}(b)). The value chosen for $\hbar \omega$ is the optimal one 
such that the condition $|E - E_{\text{exp}}| \leqslant 0.5 \, \text{MeV}$ is satisfied. The non-monotonic SRG flow of the density matrix (see Fig.~\ref{fig:densmat_n4lo_srg} of Appendix~\ref{app:densmatn4lo})
determines the dependence of $\overline{C}$ on $\lambda$. For $N \geqslant 3$, $\overline{C}$ decreases with $\lambda$
until $\lambda \sim 1.0 \, \text{fm}^{-1}$, thereby marking an overall decrease in the correlation in the HO basis. From a quantum computing point of view, this
decrease in basis correlation could be related to improved energy estimation using smaller number of qubits when SRG evolved interactions are used.

However, this 
trend does not carry over for $\lambda < 1.0 \, \text{fm}^{-1}$ and $N > 1$. For $N = 1$, $\overline{C}$ decreases monotonically for the values of $\lambda$ shown in Fig.~\ref{fig:conc}, highlighting
the suppression of the $s$-$d$ coupling with the SRG flow. The manner in which $\overline{C}$ changes with $\lambda$, for $\lambda < 1.0 \, \text{fm}^{-1}$ has no bearing on the deuteron energy estimate. We have seen that for any given $N$, the latter gets closer to the experimental value as $\lambda$ decreases. 

For the AV$_{18}$ interaction the density matrix elements are similar to those corresponding to chiral N4LO,
especially for the oscillator parameters for which the deuteron energy converges closest to the experimental value. Hence we do not present them separately.

\section{Conclusion}
\label{sect:concl}
In this paper, we have simulated the computation of the deuteron energy using Qiskit-Aer and realistic interactions. Starting from chiral N4LO and AV$_{18}$, we have evolved the interactions using the SRG formalism. In the case of exact diagonalization, as well as simulations, both in the absence and presence of gate noise, it has been shown that it is advantageous to lower the cutoff, 
since the evolved effective low-momentum interactions reduce the qubit requirement for the system at hand.

Our results with zero noise extrapolation agree both with those obtained with exact diagonalization and with noiseless simulations. This establishes the robustness of the extrapolation techniques as well as the reliability of the optimizers. Furthermore, the computed  energy agrees with the experimental value to within $1\%$. This is comparable to the results obtained using pionless EFT~\cite{Dumitrescu2018,Shehab2019,Gu2025}, and further gives significantly better realistic estimates of the deuteron energy compared to those obtained with phenomenological interactions~\cite{Siwach2021}.

We note that we get results very close to the experimental binding energy ($\sim 1.9$ to $2.2 \, \text{MeV}$)  for $\lambda \leq 1.0 \,\text{fm}^{-1}$, for 
small basis sizes ($N \leq 3$, ie., $\leq$ 6 qubits). In the two-body case, as for instance in the deuteron, the 
Wilson generator could give singular matrix elements for small $\lambda$~\cite{Arriola2016}. 
However, this neither affects the realistic value of the two-body binding energy by 
construction, nor the subsequent convergence of VQE in small oscillator spaces. 

On the other hand, it is well known that as $\lambda$ is lowered many-body forces become important. 
For example, in $^3$H, using 2N forces alone, the ground state energy differs from the experimental value: for 
$\lambda \sim 2.0 \, \text{fm}^{-1}$ the difference is $100-200 \, \text{keV}$  and for $\lambda \sim 1.0 \, \text{fm}^{-1}$ the difference is $\sim 1 \, \text{MeV}$ (see Fig.1 of~\cite{Jurgenson:2009qs}). 
However, once the leading 3N forces are included, the triton ground state energy becomes independent of $\lambda$. Similarly, in $^4$He, the ground 
state energy with two and three body interactions, differs from the experimental value by only 
about $50\, \text{keV}$ (Fig.2 
of~\cite{Jurgenson:2009qs}). However, for $^{16}$O and $^{12}$C, consistently evolving two and three-body forces to low $\lambda$ can result in the build-up of 
induced four-body forces as seen in~\cite{Roth2011} (Figs. 2 and 3). For infinite systems, such as cold neutron gas
in the crustal layers of neutron stars, the induced three-body forces grow as $\lambda$ 
decreases, but fall off to zero as the system 
becomes very dilute (Fig.6 of~\cite{Viswanathan2025}). However for symmetric matter, four-body forces become important for $\lambda \sim 1.6\, \text{fm}^{-1}$~\cite{Hebeler2021}. Therefore, for medium-mass nuclei, it is important to keep $\lambda \sim 2.0\, \text{fm}^{-1}$.  

Our work is a first step in the program for carrying out nuclear structure computations on quantum computing platforms using realistic SRG based interactions as inputs. 
It is encouraging that the SRG approach to building effective interactions, which has been employed extensively in the literature for nuclear many-body calculations, is useful in quantum computing as well.

In this work, we have used the UCC ansatz with CNOT staircases. 
We have investigated the changes in the mode entanglement (qubit entanglement) as a function of 
$\lambda$ and the oscillator parameters, using concurrence as an entanglement quantifier. There is an overall decrease in the concurrence (and therefore reduction in correlations in the HO basis) as $\lambda$ is lowered
from the bare value down to $1.0 \, \text{fm}^{-1}$. This trend in concurrence holds independent of the details of the realistic
interaction (N4LO or AV$_{18}$). The change in concurrence with $\lambda$ could probably be important in constructing other ans\"atze with
low-depth circuits that suffice for convergence to realistic values of the deuteron energy.
This would be of relevance in quantum computing for finite nuclei beyond the deuteron. The detailed behavior of concurrence for very small values of $\lambda$ ($\lambda < 1.0 \, \text{fm}^{-1}$) remains to be understood.  Work in this direction is in progress. 
Future work on entanglement between nucleons would also address the role played by the connection 
between spin and space. 

\section*{Appendix}
\appendix

\makeatletter
\@addtoreset{equation}{section}
\setcounter{equation}{0}
\renewcommand{\theequation}{\thesection\arabic{equation}}
\makeatother
\section{\label{app/two qubit} Two-qubit implementation of UCC ansatz}
\begin{figure}[h]
    \centering
    \includegraphics[width=0.5\linewidth]{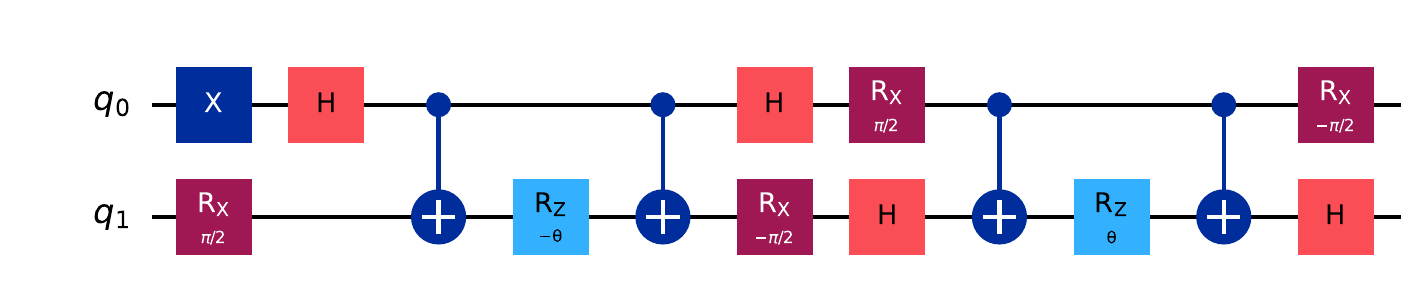}
    \caption{UCC ansatz using single excitation for the two-qubit example }
    \label{fig:circuit_example}
\end{figure}

The unitary operator $U(\boldsymbol{\theta})$, generated from the single excitation operators and the parameter vector
$\boldsymbol{\theta}$ for the one-body problem, acting on a reference state that has only the 0s mode occupied, is given by
\begin{equation}
    U(\boldsymbol{\theta}) = \exp\left[ \sum_{\alpha} {\theta}_{\alpha \,0} \left( a_\alpha^\dagger a_0 - a_0^\dagger a_\alpha \right) \right],
    \label{eq:unitary_operator}
\end{equation}
where $\alpha$ denotes the unoccupied states. 
Let us consider the two-qubit ($N = 1$) case, where there is only one parameter $\theta$. The unitary operator in Eq.~(\ref{eq:unitary_operator}) can be written 
as
\begin{equation}
    U(\theta)=e^{\theta(a_0 ^{\dagger}a_1-a_1^{\dagger} a_0)}.
\end{equation}
Using the JW mapping,
\begin{equation}
    U(\theta)=e^{-\frac{i\theta}{2}(X_0Y_1-Y_0X_1)}
\end{equation}
in terms of qubit Pauli operators. $X_0Y_1$ and $Y_0X_1$ commute. Therefore,
\begin{align}
    U(\theta)\ket{10}&=e^{-\frac{i\theta}{2}(X_0Y_1)}e^{\frac{i\theta}{2}(Y_0X_1)}\ket{10}.\\
    &=e^{-\frac{i\theta}{2}(X_0Y_1)}e^{\frac{i\theta}{2}(Y_0X_1)}X_0\ket{00}.
    \label{eq:unitary_pauli}
\end{align}
Eq.~(\ref{eq:unitary_pauli}) can be transformed into a quantum circuit using the rules (CNOT staircase construction) given in the next section and the circuit is shown in Fig.~\ref{fig:circuit_example}.
The Pauli X gate in the $0^{\rm th}$ qubit index just flips the corresponding qubit to give us the input state $\ket{10}$. For a larger number of qubits, not all the Pauli strings in the exponential commute with each other, and a single-step first-order Trotterization~\cite{Trotter1959, Llyod1996} has been employed to approximate the unitary operator to arrive at 
an ansatz similar to Eq.~(\ref{eq:unitary_pauli}). 

\section{\label{app:cnot}CNOT staircases}
\begin{figure}
        \centering
        \includegraphics[width=0.4\linewidth]{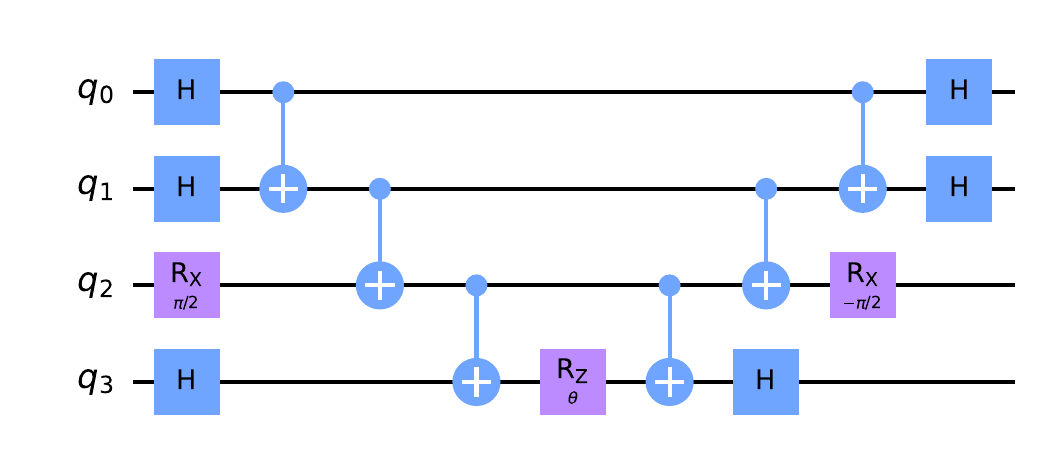}
        \caption{The circuit corresponding to $ e^{-i\theta X_0 X_1 Y_2 X_3/2}$}
        \label{fig:examplecircuit}
    \end{figure}

For operators of the form $e^{-i\theta A/2}$, where $A$ is a product of any number of single qubit Pauli matrices ($X_i$, $Y_i$, $Z_i$), a general method of CNOT staircases can be used to generate the circuit. The following identities prove to be useful:
\begin{equation}
    X_i = H_i Z_i H_i
\end{equation}
where $H_i$ is the Hadamard gate, and
\begin{equation}
    Y_i = R_{x\,i}(-\tfrac{\pi}{2}) Z_i R_{x\,i}(+\tfrac{\pi}{2}) 
\end{equation}
where $R_{x\,i}$ is the rotation about $x$-axis that acts on $i^{\rm th}$ qubit. Using the identity $e^{U^\dagger M U} = U^\dagger e^M U$, where $U$ is a unitary matrix, an operator $e^{-i X_0 X_1 Y_2 X_3 \theta/2}$ can be expressed as $\displaystyle H_0 H_1 R_{x\,2} (-\tfrac{\pi}{2})H_3  e^{-i\theta Z_0 Z_1 Z_2 Z_3/2} H_3R_{x\,2} (\tfrac{\pi}{2})H_1 H_0$. The rules for constructing the equivalent circuit using CNOT staircases can be summarized as follows:
\begin{enumerate}
    \item \label{r1} If $A=Z_0Z_1\dots Z_k \dots Z_N$, a $z$-rotation gate $R_z(\theta)$ ($R_z(\theta)=\text{cos}(\frac{\theta}{2})\mathds{1}-i\text{sin}(\frac{\theta}{2})Z$) is introduced at the highest ($N^{\rm th}$) qubit index, after sandwiching $R_z(\theta)$ between two sets of CNOT staircases that couples the $i^{\rm th}$ and the $(i-1)^{\rm th}$ qubit on either side, where $i = 0, 1, \ldots N$. 
    
    \item \label{r2} If the Pauli $X$ operator appears at the $k^{\rm th}$ position, rule~(\ref{r1}) is repeated, but now the staircases are sandwiched between two Hadamard gates at the $k^{th}$ index. 
    \item Similar to rule~(\ref{r2}) above, if the Pauli $Y$ operator appears at the $k^{\rm th}$ position, the staircases are sandwiched between an $R_x(\pi/2)$ on the left (input side) and an $R_x(-\pi/2)$ on the right (output side). 

\end{enumerate}
Hence for the example considered, the circuit is given as in Fig.~\ref{fig:examplecircuit}

\section{\label{app:chiraln4lo} Chiral N4LO interaction in momentum space}
\begin{figure*}
\centering
    \includegraphics[width=0.8\linewidth]{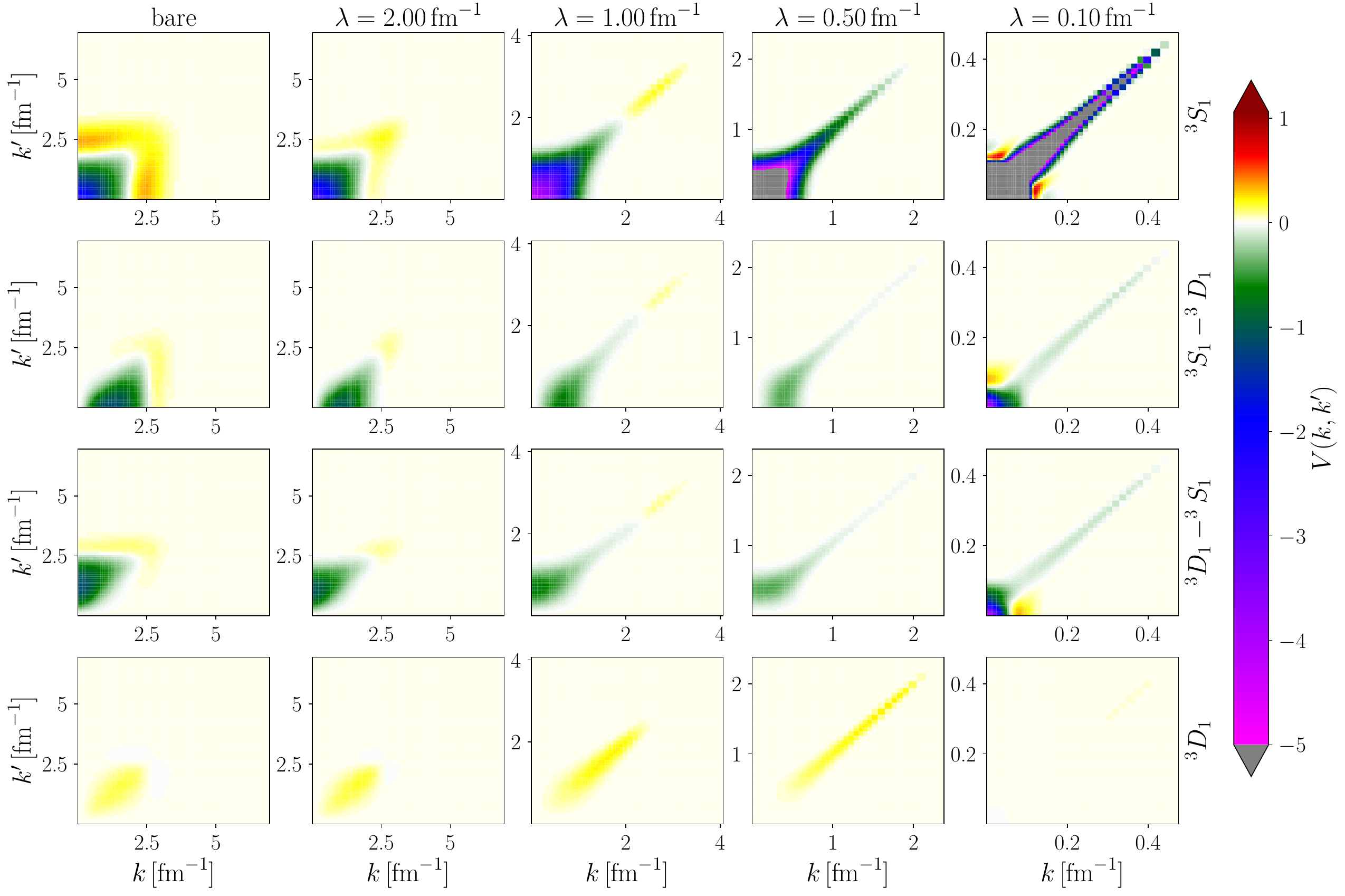}
    \caption{Band diagonal evolution of chiral N4LO in the coupled $^3S_1-^3D_1$ channel. For each partial wave, as $\lambda$ decreases, the momentum space matrix elements become increasingly band diagonal}
    \label{fig:vsrg_n4lo_3s1}
\end{figure*}
The RG evolution of the momentum space matrix elements in the deuteron channel is shown in Fig.~\ref{fig:vsrg_n4lo_3s1}. The RG flow using the Wilson's generator suppresses the far
off-diagonal matrix elements resulting in a band diagonal form as the value of $\lambda$ decreases. The strength of the tensor force that couples the $^3S_1$ and $^3D_1$ channels decreases 
with the RG flow as seen in the second and third rows of Fig.~\ref{fig:vsrg_n4lo_3s1}. Since the flow is isospectral, the two-body binding energy and 
scattering phase shifts are preserved. 

\section{Density Matrix in the HO basis for the chiral N4LO interaction}\label{app:densmatn4lo}
\begin{figure*}
    \includegraphics[width=0.9\linewidth]{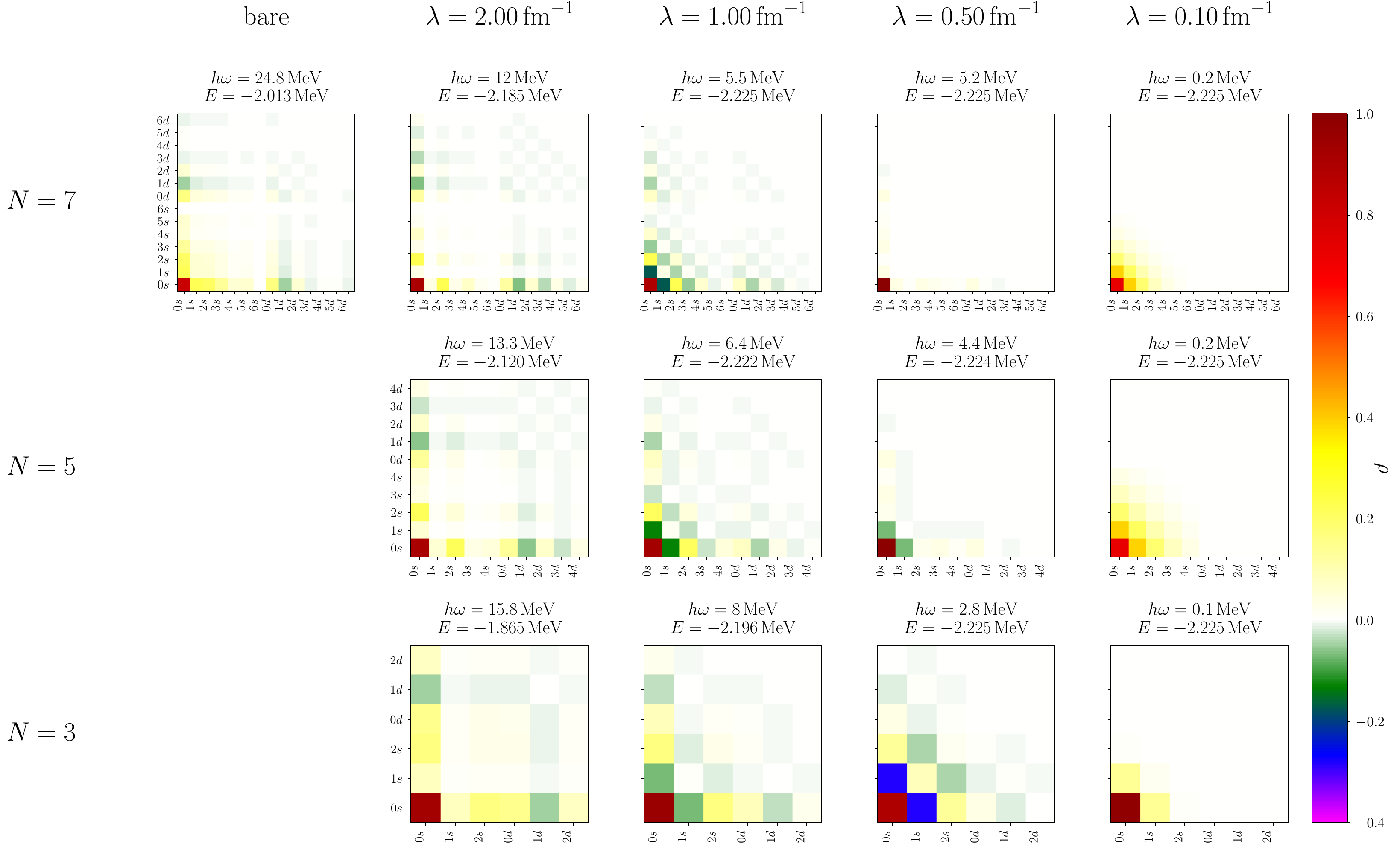}
    \caption{Density matrix in the HO basis for the chiral N4LO interaction as a function of $\lambda$ for different $N$ values for the chiral N4LO interaction. The $\hbar \omega$ for each $N$ and $\lambda$ are chosen such that $|E - E_{\text{exp}}|\leq 0.5$ MeV}
    \label{fig:densmat_n4lo_srg} 
\end{figure*}
The density matrix was computed using the deuteron wavefunction (obtained 
through exact diagonalization) in the truncated Fock basis. 
The changes in the density matrix elements 
with $N$ and $\lambda$ for values of $\hbar \omega$ chosen subject to the condition $|E - E_{\text{exp}}| \leqslant 0.5 \, \text{MeV}$, where $E$ is the optimal deuteron ground state energy, are presented in Fig.~\ref{fig:densmat_n4lo_srg}. For instance, 
when $N = 5$ or $3$, the condition imposed on the energy is satisfied only for $\lambda \lesssim 2.0 \, \text{fm}^{-1}$ in
contrast to $N = 7$. For any $N$, as $\lambda$ decreases, the non-zero matrix elements of the density operator shift 
to lower oscillator modes. Further, the strength of the lowest mode (0s) increases as $\lambda$ and $N$ decrease. 
The coupling between the $s$ and $d$ modes for any $n$ becomes negligible as $\lambda$ decreases reflecting the weakening of the tensor force. The details of the
changes in the density matrix elements as a function of $\lambda$ are presumably due to the intricacies of the SRG evolution 
and the choice of the generator~\cite{Arriola2016}.  

\section{\label{app:mode_ent}Mode entanglement}
Let the deuteron state be represented by $\ket{\Psi}$ and the mode of interest by $\ket{\tilde{n},\tilde{l}}$. We study the entanglement of this mode (subsystem 1) with the rest of the modes (subsystem 2).  In terms of the superposition coefficients 
$\braket{n, l|\Psi}$, the total density matrix is,
\begin{equation}\label{mode_total_density_mat}
    \rho=\ket{\Psi}\bra{\Psi}=\sum_{n, l, n', l'}\braket{n, l|\Psi}\braket{\Psi|n', l'} \ket{n, l}\bra{n', l'}.
\end{equation}
In the occupation number representation, $\ket{\tilde{n}, \tilde{l}}$ will be represented as $\ket{10000\dots}$, where for notational convenience, the first label is reserved for the mode in consideration. The rest of the labels represent the remaining modes. We denote a general ket in this representation by $\ket{xy}$ where $x$ can be 0 or 1 and $y$ is a binary string of 0s and 1s. By construction, if $x=1$, then $y$ must be a string of all zeros. The expansion coefficients of the deuteron state in this basis are 
written as $C_{xy} = \braket{xy|\Psi}$. In this representation, the density matrix in Eq.(\ref{mode_total_density_mat}) can be rewritten as,
\begin{align}
    \rho=\sum_{x,y, x', y'}^{} C_{xy}C^{*}_{x'y'}\ket{xy}\bra{y'x'}.
    \label{mode_total_density_mat_xy}
\end{align}
The single mode reduced density matrix $\tilde{\rho}$, found by tracing over the rest of the modes (all possible combinations of $y$), is given by,

\begin{align}
  \tilde{\rho}&=\sum_{y_1}\braket{y_1|\Psi}\braket{\Psi|y_1} \nonumber \\
  &= \sum_{y_1}\sum_{x,y, x', y'} C_{xy}C^{*}_{x', y'} \delta_{y_1y}\delta_{y_1y'}\ket{x}\bra{x} \nonumber \\
  & =\sum_{y_1}\sum_{x, x'} C_{xy_1} C^{*}_{x'y_1} \ket{x}\bra{x'}.
\end{align}

It can be shown that in the matrix representation, $\tilde{\rho}$ is a $2 \times 2$ diagonal matrix with diagonal entries given in terms of the superposition coefficients as $|\braket{\tilde{n}, \tilde{l}|\Psi}|^{2}$ and  $1-|\braket{\tilde{n}, \tilde{l}|\Psi}|^{2}$. Therefore, for the single mode case, the concurrence, defined as $C=\sqrt{2(1-\mathrm{Tr}(\tilde{\rho}^{2}))}$ becomes
\begin{align}
    C=\sqrt{2(1-(|\braket{\tilde{n}, \tilde{l}|\Psi}|^{4}+(1-|\braket{\tilde{n}, \tilde{l}|\Psi}|^{2})^{2}))}.
\end{align}

\ack{
We thank V. Palaniappan for the SRG codes that were used to generate the two-body matrix elements for this work. We thank the referee for bringing to our notice reference~\cite{Roth2011} of the manuscript. We acknowledge partial support through funds from the Center for Quantum Information, Computing and Communication (CQuICC), IIT Madras. SL and VB thank the Department of Physics, IIT Madras for infrastructural support.}

\printbibliography

\end{document}